\documentclass[10pt,prl,twocolumn,showpacs]{revtex4-1}
\pdfoutput=1
\usepackage{graphicx,amsmath,amssymb,amsthm}
\usepackage{hyperref}
\usepackage{xcolor}
\usepackage{ulem}
\usepackage{thmtools, thm-restate}

\begin{document}
\newtheorem*{conjecture1}{Distance Conjecture:}
\newtheorem*{conjecture2}{De-Sitter Swampland Conjecture:}
\def\nn{\nonumber}
\def\kc#1{\left(#1\right)}
\def\kd#1{\left[#1\right]}
\def\ke#1{\left\{#1\right\}}
\newcommand\beq{\begin{equation}}
\newcommand\eeq{\end{equation}}
\renewcommand{\Re}{\mathop{\mathrm{Re}}}
\renewcommand{\Im}{\mathop{\mathrm{Im}}}
\renewcommand{\b}[1]{\mathbf{#1}}
\renewcommand{\c}[1]{\mathcal{#1}}
\renewcommand{\u}{\uparrow}
\renewcommand{\d}{\downarrow}
\newcommand{\be}{\begin{equation}}
\newcommand{\ee}{\end{equation}}
\newcommand{\bsigma}{\boldsymbol{\sigma}}
\newcommand{\blambda}{\boldsymbol{\lambda}}
\newcommand{\Tr}{\mathop{\mathrm{Tr}}}
\newcommand{\sgn}{\mathop{\mathrm{sgn}}}
\newcommand{\sech}{\mathop{\mathrm{sech}}}
\newcommand{\diag}{\mathop{\mathrm{diag}}}
\newcommand{\Pf}{\mathop{\mathrm{Pf}}}
\newcommand{\half}{{\textstyle\frac{1}{2}}}
\newcommand{\sh}{{\textstyle{\frac{1}{2}}}}
\newcommand{\ish}{{\textstyle{\frac{i}{2}}}}
\newcommand{\thf}{{\textstyle{\frac{3}{2}}}}
\newcommand{\SUN}{SU(\mathcal{N})}
\newcommand{\N}{\mathcal{N}}

\title{A Potential Mechanism for Inflation from Swampland Conjectures}

\author{Hao Geng}
\affiliation{Department of Physics, University of Washington, Seattle, WA, 98195-1560, USA}
%\author{Xiaojun Yao}
%\affiliation{Department of Physics, Center for Theoretical Physics, Massachusetts Institute of Technology, Cambridge, MA, 02139, USA}
\date\today

\begin{abstract}
Inflation is the currently accepted paradigm for the beginnings of the Universe. To explain the observed almost scale invariant spectrum of density perturbations with only a slight spectral tilt, inflation must have been ``slow roll", that is with a potential with sufficiently small slope. While the origin of inflationary structure is intrinsically quantum mechanical, gravity gets treated semiclassically within inflationary models. Recent work, in terms of the so called de-Sitter swampland conjecture, has called into question whether slow roll inflation is consistent with a complete theory of quantum gravity in the presence of a positive vacuum energy density, which is a key ingredient in the inflationary paradigm. In this work, we show that, in fact, if we understand this conjecture correctly and with another swampland conjecture, the so-called distance conjecture, involved we get a potential mechanism for slow roll inflation and we argue that here fine-tuning is not a technical problem.
\end{abstract}

\pacs{04.20.Cv,
%Fundamental problems and general formalism
04.60.Bc,
%Phenomenology of quantum gravity
98.80.Qc
% Quantum cosmology
%%[[fix these]]
}
\maketitle

{\bf Introduction:}
Inflation\cite{Starobinsky:1980te,Guth:1980zm,Linde:1981mu,Albrecht:1982wi}, a period of exponential expansion driven by a scalar field called the inflaton, is the current paradigm for the origin of our universe, and in particular for the seeds of structure within it that eventually grew into clusters, galaxies, stars and planets. Inflation speaks to several theoretical puzzles: why is the universe spatially flat? Why does the universe look homogeneous and isotropic and the largest scales, seemingly implying equilibration of causally disconnected regions? But most importantly, inflationary models make clear predictions \cite{Press:1980zz,Hawking:1982cz,Starobinsky:1982ee,Guth:1982ec,Bardeen:1983qw,Mukhanov:1981xt} for the spectrum of density fluctuations in the early universe. Quantum fluctuations get stretched to macroscopic scales by the inflationary expansion, and gravity does the rest. Overdense regions attract more matter which eventually collapses under its own attractions and leads to all the visible structure in the universe. The almost scale invariant spectrum implied by inflation is in stunning agreement with observations of the cosmic microwave background \cite{Akrami:2018odb}. 

Most inflationary models require the evolution of the scalar field $\phi$ to be ``slow roll". That is the scalar field is slowly rolling down a relatively shallow scalar potential. Technically this is incorporated in the so called slow roll parameters which characterize the slope and curvature of the potential $V(\phi)$:
\beq
\epsilon \equiv \frac{M_P^2}{16 \pi} \left ( \frac{ V'}{V} \right )^2, \quad
\eta \equiv \frac{M_P^2}{8 \pi} \left ( \frac{ V''}{V} \right ).\label{eq:slowroll}
\eeq
Slow roll inflation corresponds to choosing $\epsilon, \eta \ll 1$. In all simple inflationary models slow roll is absolutely necessary for two reasons. For one, the number of e-folds, that is the logarithm of the final over initial scale factor of the universe, is proportional to $\epsilon^{-1/2}$. To get a sufficient number of e-folds to solve the horizon and flatness problems to begin with one needs either a small epsilon or large changes in the scalar field value compared to the Planck scale. More quantitatively, $\epsilon$ and $\eta$ lead to deviations from a scale invariant power spectrum of the fluctuations and so the experimentally observed almost scale invariant spectrum directly implies upper bounds on $\epsilon$ and $\eta$. This is quantified via the so called spectral tilt $n_s-1$. $n_s=1$ corresponds to a completely scale invariant power spectrum, whereas $n_s$ that deviates from that value tell us that the power spectrum has a small variation with $k$. In a slow roll inflation model one can find that \cite{Stewart:1993bc}
\beq
n_s -1 = - 6 \epsilon + 2 \eta.
\eeq
The degeneracy between $\epsilon$ and $\eta$ can be broken by considering higher order variations in $k$. Using the 2018 data from the Planck mission on the observed power spectrum in the cosmic microwave background (CMB) radiation one can derive an observational {\it upper} bound on $\epsilon$ of
\beq
\epsilon < 0.0063\label{eq:exp}
\eeq
at the 95 \% confidence level \cite{Akrami:2018odb}. However, potentials support the smallness of $\epsilon$ are generally unnatural with the same reason as the long-standing issue of Higgs mass in the Standard Model of particle physics. Furthermore, inflation also has other deep theoretical shortcomings, see e.g. \cite{Ijjas:2014nta}, that question some of its phenomenological successes. Those problems are almost all about the unknown underlying origin of inflation namely looking at it from an effective field theory point of view without addressing its ultra-violet(UV) completion. 

This urges a better understanding of reality from an underlying UV complete point of view. String theory is the only known consistent framework of a UV complete unifying theory and so a lot of string theory motivated conjectures, called swampland conjectures, distinguishing those effective field theories able to be embedded in string theory to those cannot, are proposed \cite{Ooguri:2006in,Obied:2018sgi,Ooguri:2018wrx}. Among these conjectures, a so called de-Sitter swampland conjecture \cite{Obied:2018sgi} and its refined version \cite{Garg:2018reu} strongly denied the existence of slow roll inflation \cite{Garg:2018reu,Garg:2018zdg,Kinney:2018nny}. However, the potential addressed by those authors should be clarified. As from several explicit examples provided by \cite{Obied:2018sgi} from string theory compactifications, we here claim that the potential addressed by them is the geometric potential which is part of Einstein-Hilbert action (possibly with cosmological constant) from a higher dimensional decompactified point of view. Hence these potentials did not fully address the interaction between the scalar (modulus) field and hidden sectors emerging from the geometry of string theory.

In this work, we will try to study the whole quantum interaction effect of string theory compactification onto inflation, due to the so called distance conjecture\cite{Ooguri:2018wrx}, and see that in a complete description of inflation, with the de-Sitter conjecture and the quantum interaction effect, slow roll inflation actually has a potential mechanism to be engineered from string theory\footnote{Other recent studies for swampland conjectures motivated inflation mechanisms can be found in for example \cite{Brahma:2018hrd,Brahma:2019mdd,Brahma:2019unn}}. Therefore, the existing problems \cite{Ijjas:2014nta}  of inflation might have a unified answer in this complete description. We'll discuss issues about fine-tuning at the end. It deserves to be mentioned that several effects that can flatten inflaton potentials are studied in a nice paper\cite{Dong:2010in} and a recent study of the violation of slow-roll conditions for inflaton by coupling it to gravity is \cite{Takahashi:2020car}.

{\bf Setup and notation:} We will consider a scalar field constructed from a consistent quantum gravitational theory, for example string theory. To address the complete dynamics of this scalar field including potential interactions with other sectors, we will use the distance conjecture \cite{Ooguri:2006in}.

\begin{conjecture1}
As the modulus is displaced a distance of the order of a Planck scale in moduli space, there would be a tower of light states that emerges with masses exponentially suppressed.
\end{conjecture1}
In this statement, moudlus refers to a scalar field $\phi$ parametrizing the compactification or roughly speaking the size of an extra dimension and, in our context, we will take it to be the inflaton. The emergent modes are usually the so called Klauza-Klein (KK) modes and winding modes from string theory compactification\cite{Johnson:2003gi,Polchinski:1998rq} (other examples can be found at \cite{Ooguri:2006in}). Moreover, the KK modes will be heavy for tiny extra dimensions and will be light for large extra dimensions and the winding modes will behave in a opposite way. Without loss of generality, we will first consider a single copy of the emergent mode described as a scalar field $\Phi$ with mass square
\beq
   m^{2}(\phi)=m^{2}_{0}e^{-c\frac{\phi}{M_{P}}}\label{eq:mass}
\eeq
where $M_{P}$ is the Planck scale, $c$ is an order one constant and $m^{2}_{0}$ is defined in a way that $\phi$ starts at 0 \footnote{Here we should be careful that as the statement goes this exponential form of the modulus dependence of mass is generally expected when $\phi\sim M_{P}$ where the mode is not guaranteed to be heavy. Even though in some simple string theory compactifications, for example compactifying a dimension on a circle, this dependence is exact, we would expect that the massof the emergent mode will have a more complicated dependence on $\phi$ in a generic compactification. However, we believe the effect that integrating out this heavy mode reduces the slow roll parameters (\ref{epsilon}) and (\ref{eta}) would persist as a generic mechanism. Hence we still choose the exponential form of $m^{2}(\phi)$}.

From now on we will call the modulus field $\phi$ as inflaton. We will consider the following Lagrangian density
\begin{equation}
    \mathcal{L}=\frac{1}{2}\partial_{\mu}\phi\partial^{\mu}\phi-V(\phi)+\frac{1}{2}\partial_{\mu}\Phi\partial^{\mu}\Phi- \frac{1}{2}m^{2}(\phi)\Phi^{2}
\end{equation}
where $\Phi$ is the emergent mode from the distance conjecture and $V(\phi)$ is the potential of inflaton from string theory compactification. This potential satisfies the following conjecture \cite{Obied:2018sgi}.
\begin{conjecture2}
 Scalar field potentials arising from a consistent quantum gravitational theory should satisfy
 \begin{equation}
     |\nabla{V}|\geq \tilde{c} \text{ }V
 \end{equation}
 where $\tilde{c}$ is of order one in Planck unit.
\end{conjecture2}
As it frequently arises in string theory compactifications, we will take the exponential form of the inflaton potential
\beq
   V(\phi)=Be^{-b\frac{\phi}{M_{P}}}, \text{ }b\sim O(1).
\eeq
The comment on inflation and current experiment on the CMB power spectrum \eqref{eq:exp} from this conjecture is that it is in tension with inflation but not with \eqref{eq:exp}. Because as we discussed in the introduction that to resolve the horizon and flatness problems using inflation we need slow roll $\epsilon, \eta \ll 1$. And to satisfy the current experimental upper bound we only have to require that $b<0.564$ which is not necessarily in conflict with $b\sim\mathcal{O}(1)$.

{\bf Integrating out the emergent modes:}
Since we want to identify a mechanism for slow roll inflation, we'll assume $\phi$ is slow rolling and check that if there is a mechanism supporting this assumption given that the dS swampland conjecture is satisfied. We want to consider the complete dynamics of $\phi$ including its interaction with the emergent mode and hence we will integrate out the emergent mode. Also, we assume that the characteristic mass scale $m_{0}$ of the emergent modes is heavier than that of $\phi$. As a prelude, we'll exactly integrate it out and focus on the resulting dynamics for the inflation. The result is that we have a quantum effective theory of inflaton organized systematically by the small expansion parameter $\frac{\partial^{2}}{m_{0}^{2}}$ and $\mathcal{O}(x)=e^{-c\frac{\phi}{M_{P}}}-1$ and they are small because of the slow roll of $\phi$. Our resulting effective potential is
\beq
\begin{split}
    V_{\text{eff}}=&Be^{-b\frac{\phi}{M_{P}}}+\frac{m_{0}^{4}}{32\pi^{2}}\Big[\log(\frac{\tilde{\mu}^{2}}{m_{0}^{2}e^{\frac{1}{2}}})(1-e^{-2\frac{c\phi}{M_{pl}}})+\\
    &\frac{1}{2}\log(\frac{\tilde{\mu}^{2}e^{\frac{3}{2}}}{m_{0}^{2}})(1-e^{-2\frac{c\phi}{M_{pl}}})^{2}+\frac{c\phi}{M_{pl}}e^{-4\frac{c\phi}{M_{pl}}}\Big]\label{eq:Veff}
  \end{split}
\eeq
where $\tilde{\mu}$ is the cutoff scale beyond which a complementary light state will take over. In the language of string theory we can take $\tilde{\mu}$ to be the self-dual scale under T-duality \cite{Polchinski:1998rq,Johnson:2003gi,Tseytlin:1991xk}. Details of how to derive this exact effective potential can be found in the appendix.

\begin{figure}
\centering
\includegraphics[width=8.1cm]{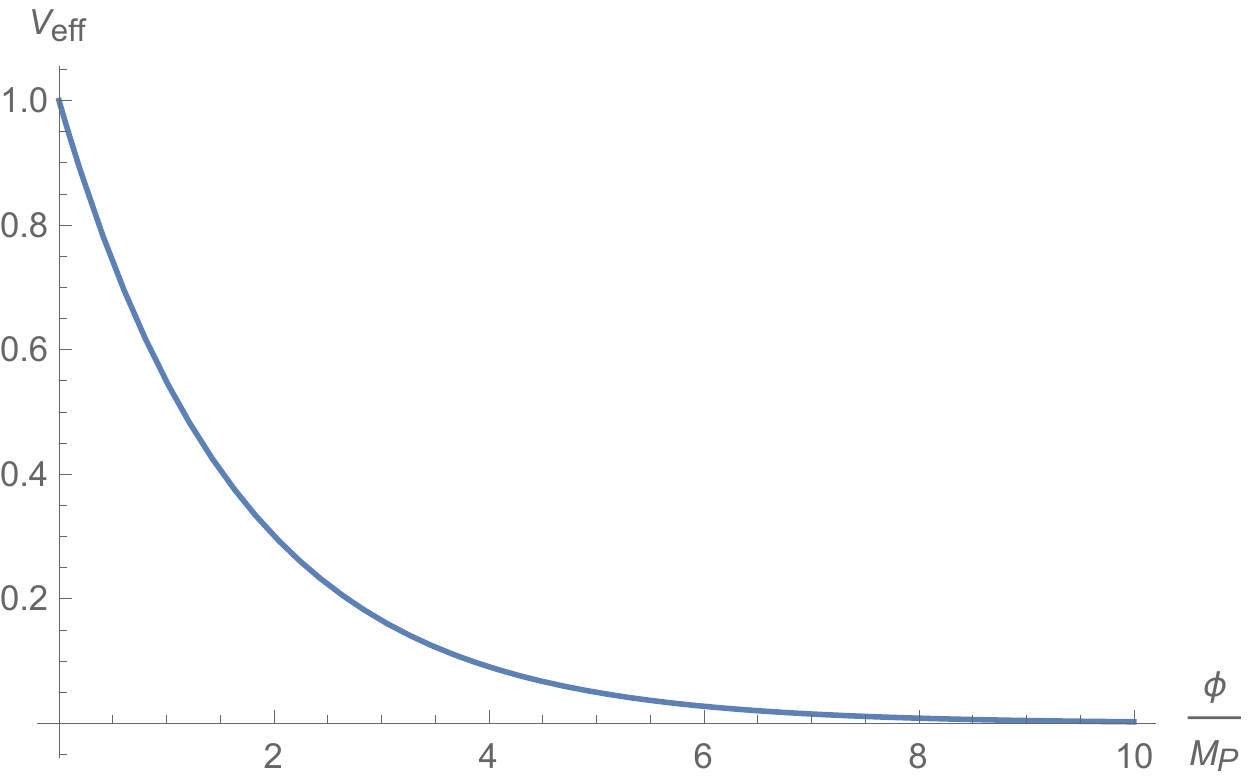}
\includegraphics[width=8.1cm]{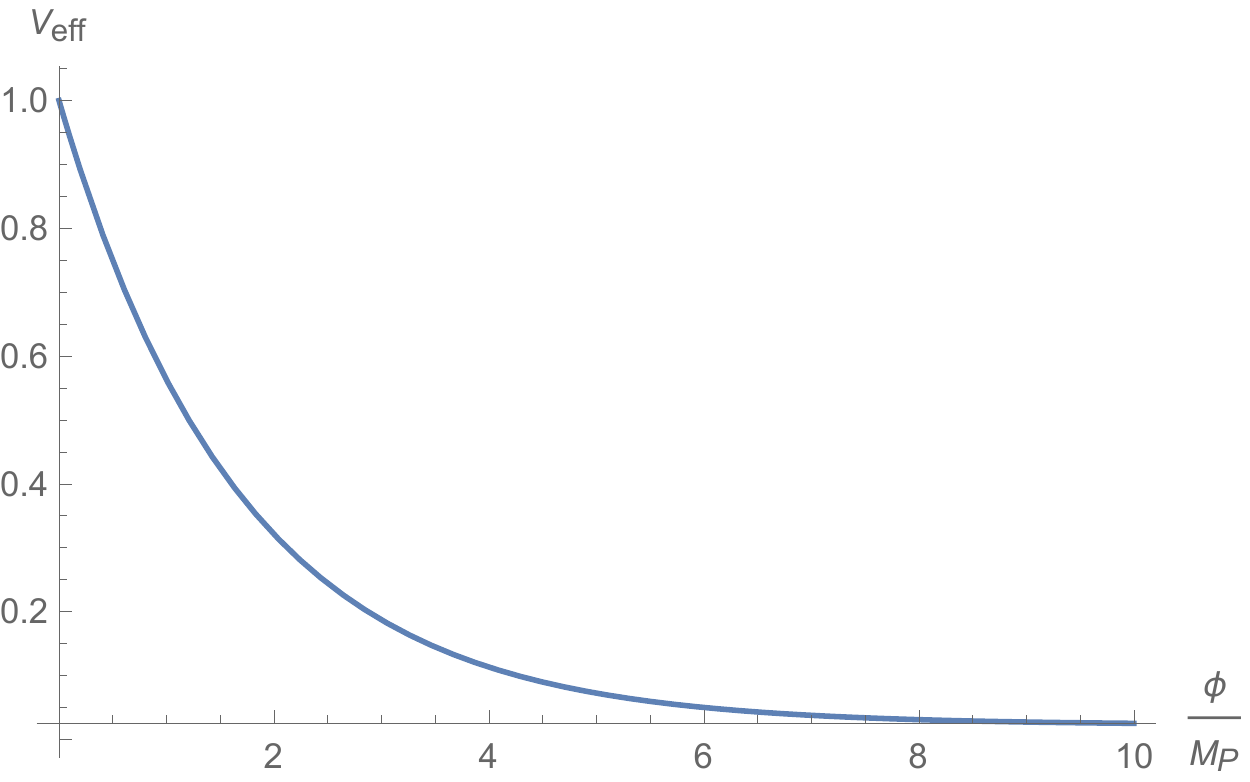}
\includegraphics[width=8.1cm]{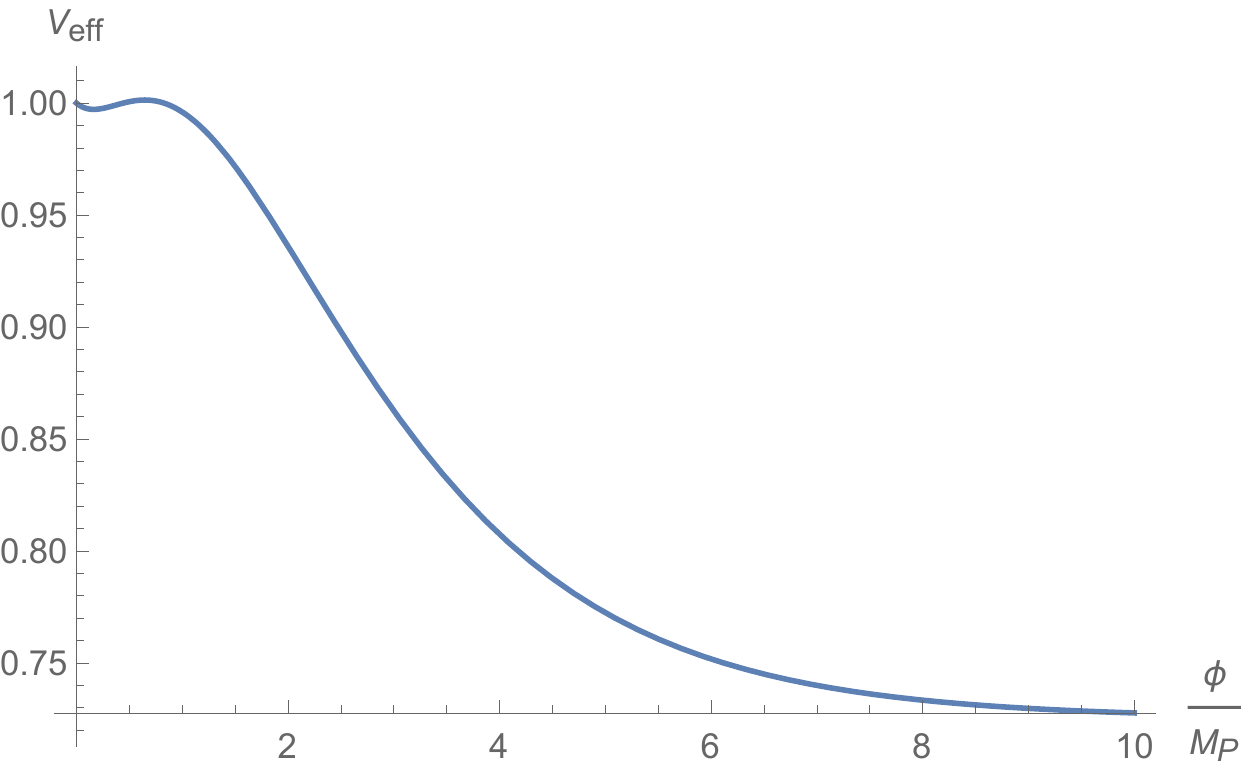}
\includegraphics[width=8.1cm]{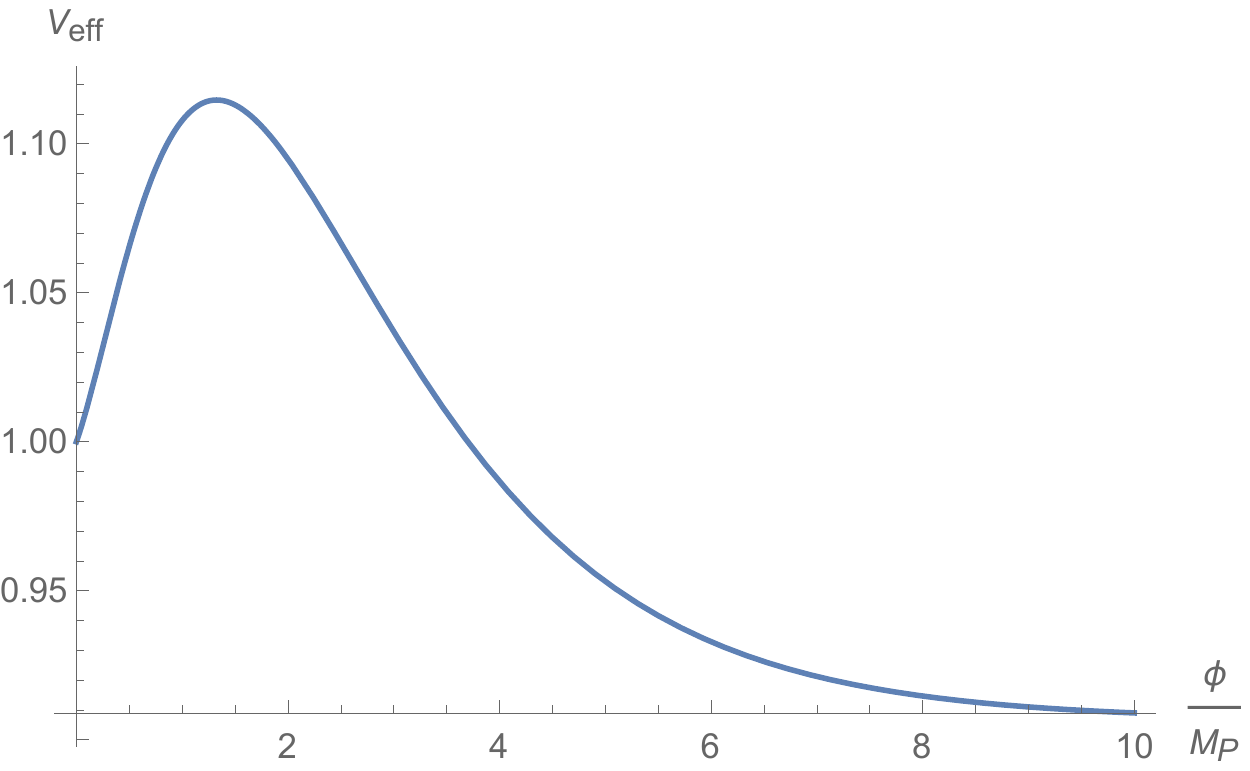}
\caption{For demonstration we take $b=c=0.6$, $B=m_{0}=1$, $\frac{\tilde{\mu}}{m_{0}}=10$ and we draw $V_{\text{eff}}$ for $N=0,1,32 \text{ and }40$. We can see that for $N=32$ the potential has an initially very flat platform which lasts for approximately a Planck scale of the value of $\phi$ and after that the potential has a steep valley which is exactly the requirement of inflation and its termination \cite{Baumann:2009ds}.}
\label{pic:inflation}
\end{figure}

{\bf Slow roll parameters:}
Slow roll parameters are defined in \eqref{eq:slowroll}. In this section and next we will see that slow roll can be supported by our effective potential \eqref{eq:Veff}. We will use the fact that among the slow roll regime, $\phi$ starting from the origin, $e^{-c\frac{\phi}{M_{p}}}\sim1$. The slow roll parameters are given approximately by
\beq
 \epsilon\approx\frac{b^{2}}{16\pi}\Big[1-\frac{m_{0}^{4}c}{16\pi^{2}Bb}\log(\frac{\tilde{\mu}^{2}}{m_{0}^{2}})\Big]^{2}\label{epsilon},
\eeq
\beq
  \eta\approx\frac{b^{2}}{8\pi}\Big[1+\frac{m_{0}^{4}c^{3}}{4\pi^{2}Bb^{2}}\frac{\phi}{M_{P}}(1-2\log(\frac{\tilde{\mu}^{2}}{m_{0}^{2}}))\Big]\label{eta}.
\eeq
It deserves to be noticed that here we only considered one emergent mode. However as we know from the distance conjecture that near the boundary of the moduli space there will be a tower of light modes emerging. Hence even though we might not be very close to the boundary we should consider the multiple emergent modes effect. The effect is very simple if the only couplings are the modulus-dependent masses. For the sake of convenience, we assume that there are $N$ emergent modes with the same masses as (\ref{eq:mass}) then the slow roll parameters are simply
\beq
 \epsilon\approx\frac{b^{2}}{16\pi}\Big[1-\frac{Nm_{0}^{4}c}{16\pi^{2}Bb}\log(\frac{\tilde{\mu}^{2}}{m_{0}^{2}})\Big]^{2}\label{epsilon1},
\eeq
\beq
  \eta\approx\frac{b^{2}}{8\pi}\Big[1+\frac{Nm_{0}^{4}c^{3}}{4\pi^{2}Bb^{2}}\frac{\phi}{M_{P}}(1-2\log(\frac{\tilde{\mu}^{2}}{m_{0}^{2}}))\Big]\label{eta1}.
\eeq
Interestingly, with a moderate number of emergent modes these slow roll parameters are further reduced and after a Planck scale of the inflaton's displacement there is a steep valley of the inflaton potential (see Fig.\ref{pic:inflation}). If we take the values in Fig.\ref{pic:inflation} and $N=32$, we find that near $\phi\sim0.4 M_{P}$ the slow roll parameters and the inflationary observables $n_{s},\text{ }r$- spectral tilt and the tensor to scalar ratio have the values
\begin{equation}
\begin{split}
    \epsilon&=3.319\times10^{-6}, \eta=-5.810\times 10^{-4}\\n_{s}&=0.9988, r=10\epsilon=3.319\times10^{-5}.
\end{split}
\end{equation}

{\bf Discussion and future remarks:}
From the expressions of slow roll parameters \eqref{epsilon} and \eqref{eta}, we see that the cutoff dependence is logarithmic and hence our theory is not fine-tuned from the field theory point of view. However, in some sense we do need "fine-tuning" by adjusting several parameters for example $B,b$, $c$, $m_{0}$ and the self-dual scale $\tilde{\mu}$  but they can be engineered in string theory compactification. As a result, from this UV-complete description we saw that fine-tuning is not a technical problem for our mechanism of inflation but just a way to look at the huge landscape of string theory vacua. A possible question is that for string theory compactification we always have KK modes and winding modes the masses of which behave in opposite ways as the size of the extra dimension changes so in addition to \eqref{eq:mass} there should be other modes with exponentially growing mass. And these heavy modes would have opposite contributions to slow roll parameters from the light modes. However, the point is that distance conjecture is true only for trans-Planckian moduli distance namely when the extra dimension parametrized by $\phi$ has either very small radius or very large radius. In the intermediate regime these two effects are competing and hence cancel each other in the loop. And one of them will dominate over the other for trans-Planckian moduli distances which is the case we are considering. Another possible question about how large $\frac{\tilde{\mu}}{m_{0}}$ is because it determines whether the logarithmic terms are positive or negative and how positive or negative they are. We could say for sure that $m_{0}$ is lower than the self-dual scale $\tilde{\mu}$ because as we said before that if it goes over $\tilde{\mu}$ then the complementary sector will be light and take over in the low energy effective theory. As a result, it is for sure that the logarithmic terms are positive and hence $\epsilon$ is reduced. Now the question is if it is possible that the logarithmic terms are bigger than one and therefore $\eta$ will also be reduced. For this question we did not see any obstacle why this is not possible. And this the reason why we think that our mechanism is a potential mechanism but not a literal mechanism for inflation. If it is the case that the logarithm term must be smaller than one then we might have to consider the refined de-Sitter swampland conjecture \cite{Ooguri:2018wrx}. A potential short-coming of our mechanism is that we did not consider the field theory in a curved spacetime. Moreover, our effective action can be used to understand the relationship between infinite distance and an infinite tower of emergent modes from the point of effective field theory as suggested in \cite{Ooguri:2006in} in a closed form to all orders in perturbation theory. We will leave these questions for future studies.

The last thing we would like to address is that the geometry of string theory might change our way of thinking about low energy physics from the traditional quantum field theory point of view (more examples of this view point can be reached at e.g. \cite{Geng:2019zsx,Reece:2018zvv}). 

{\bf Acknowledgements:} I appreciate Gily Elor, Andreas Karch, Matthew McQuinn, Cumrun Vafa and Xiaojun Yao for useful discussions. Moreover, I specially thank an anonymous reviewer for providing enormous valuable suggestions helping this letter to be published. This work is supported in part by a grant from the Simons Foundation (651440, AK). I am very grateful to my parents and recommenders.
\appendix
\section{Details on the effective action}
In this section we will show details of our effective theory which is one loop exact. Our effective theory is controlled by two small parameters-$\frac{\partial^{2}}{m_{0}^{2}}$ and $\mathcal{O}(k)=\widetilde{(e^{-2\frac{c\phi}{M_{pl}}}-1)}(-k)$. The first one is small because we are in slow roll regime which is relevant for inflation and the second one is small in the sense that as an operator its expectation value is small because we defined our theory near $\phi=0$ and slow roll.

The effective action of $\phi$ reads
\begin{equation}
\begin{split}
    S_{eff}&=S_{free}+\frac{i}{2}\Tr \log[\partial^2+m_{0}^{2}e^{-2\frac{c}{M_{pl}}\phi}]\\
    &=S_{free}+\frac{i}{2}\Tr\log[(\partial^{2}+m_{0}^{2})(1+\frac{m_{0}^{2}(e^{-2\frac{c\phi}{M_{pl}}}-1)}{\partial^{2}+m_{0}^{2}})]\\
    &\equiv S_{free}+ \frac{i}{2}\Tr\log[1+\frac{m_{0}^{2}(e^{-2\frac{c\phi}{M_{pl}}}-1)}{\partial^{2}+m_{0}^{2}}]\\
    &=S_{free}+ \frac{i}{2}\Tr\log[1+\frac{1}{1+\frac{\partial^{2}}{m_{0}^{2}}}(e^{-2\frac{c\phi}{M_{pl}}}-1)]\\
    &=S_{free}+ \frac{i}{2}\Tr\log[1+\frac{1}{1+\frac{\partial^{2}}{m_{0}^{2}}}\mathcal{O}(x)]
    \end{split}
\end{equation}
where the $\log(\partial^{2}+m_{0}^{2})$ in the second step only gives an overall normalization factor of the partition function and hence is dropped.

\begin{figure}
\centering
\includegraphics[width=0.6\linewidth]{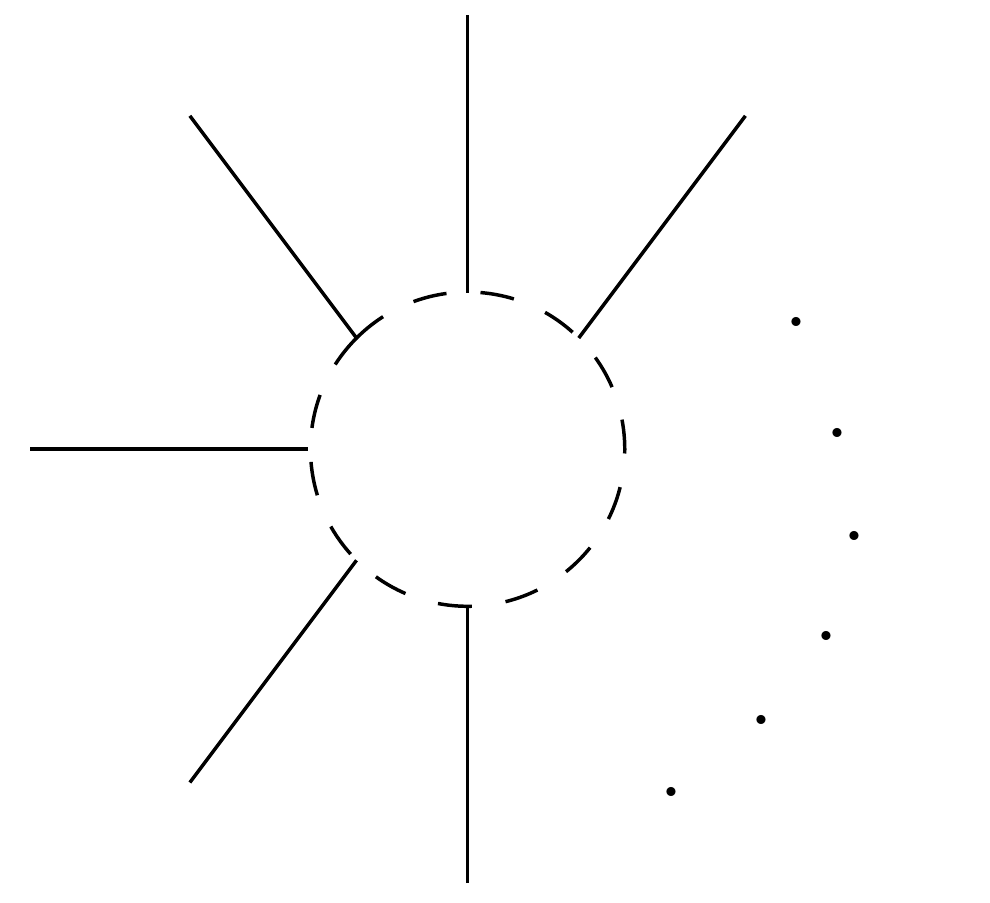}
\caption{Relevant Feynman diagrams.}
\label{pic:loop}
\end{figure}

From here we see that we can systematically reorganize the effective action using the two small parameters mentioned at the beginning of this section. Furthermore, this is the exact effective action of the inflaton (the only Feynman diagrams are of type Fig.\ref{pic:loop}) defined at energy scales below $m_{0}$ and the regime that $\phi$ is slowly rolling. The effective potential, i.e. the non-derivative part of the effective action is found to be \eqref{eq:Veff}. $\tilde{\mu}$ is the cutoff scale of the emergent modes inside the loop and it is the scale beyond which the complementary emergent mode will take over. For example, if it is a KK mode then the complementary mode is a winding mode and vice versa.

The first term in $\Tr$ expansion can be worked out as following:
\begin{equation}
    \begin{split}
        \Gamma^{(1)}&=\frac{i}{2}\Tr[\frac{m_{0}^{2}}{\partial^{2}+m_{0}^{2}}\mathcal{O}(x)]\\&=\frac{i}{2}\int d^{4}x\mathcal{O}(x)\int\frac{d^{4}k}{(2\pi)^{4}}\frac{m_{0}^{2}}{-k^{2}+m_{0}^{2}}\\
        &=-\frac{1}{32\pi^{2}}\int d^{4}x\mathcal{O}(x)(m_{0})^{4}\log(\frac{\tilde{\mu}^{2}}{m_{0}^{2}}).
    \end{split}
\end{equation}
Derivative starts to appear at the second term
\begin{equation}
    \begin{split}
    \Gamma^{(2)}=&-\frac{i}{4}\Tr[\frac{m_{0}^{2}}{\partial^{2}+m_{0}^{2}}\mathcal{O}(x)\frac{m_{0}^{2}}{\partial^{2}+m_{0}^{2}}\mathcal{O}(x)]\\
    =&-\frac{im_{0}^{4}}{4}\int\frac{d^{4}k}{(2\pi)^{4}}\frac{d^{4}p}{(2\pi)^{4}}\frac{1}{-p^{2}+m_{0}^{2}}\frac{1}{-(p+k)^{2}+m_{0}^{2}}\\&\times\widetilde{\mathcal{O}}(k)\widetilde{\mathcal{O}}(-k)\\
    =&\frac{m_{0}^{4}}{64\pi^{2}}\int\frac{d^{4}k}{(2\pi)^{4}}\int_{0}^{1}d\eta\log(\frac{\tilde{\mu}^{2}}{m_{0}^{2}-\eta(1-\eta)k^{2}})\times\widetilde{\mathcal{O}}(k)\widetilde{\mathcal{O}}(-k)\\
    =&\frac{m_{0}^{4}}{64\pi^{2}}\int d^{4}x\int_{0}^{1}d\eta\mathcal{O}(x)\log(\frac{\tilde{\mu}^{2}}{m_{0}^{2}+\eta(1-\eta)\partial^{2}})\mathcal{O}(x)\\
    =&\frac{m_{0}^{4}}{64\pi^{2}}\int d^{4}x\int_{0}^{1}d\eta\mathcal{O}(x)\times[\log(\frac{\tilde{\mu}^{2}}{m_{0}^{2}})+O(\frac{\partial^{2}}{m_{0}^{2}})]\mathcal{O}(x)
    \end{split}
\end{equation}
where $\eta$ is the Feynman parameter and for the effective potential we can drop the derivative terms. Furthermore, we can work out the general term for $n > 2$ as
\begin{equation}
\begin{split}
    \Gamma^{(n)}=&\frac{i(-1)^{n-1}}{2n}\Tr\Big\{\big[\frac{m_{0}^{2}}{\partial^{2}+m_{0}^{2}}O\big]^{n}\Big\}\\=&-\frac{im_{0}^{2n}}{2n}\int\frac{d^{4}p_{1}}{(2\pi)^{4}}\cdots\frac{d^{4}p_{n}}{(2\pi)^{4}}\\&\times\frac{\widetilde{O}(p_{1}-p_{2})\widetilde{O}(p_{2}-p_{1})\cdots\widetilde{O}(p_{n}-p_{n-1})}{(p_{1}^{2}-m_{0}^{2})(p_{2}^{2}-m_{0}^{2})\cdots(p_{n}^{2}-m_{0}^{2})}\\=&-\frac{im_{0}^{2n}}{2n}\int\frac{d^{4}k_{1}}{(2\pi)^{4}}\cdots\frac{d^{4}k_{n}}{(2\pi)^{4}}\widetilde{O}(k_{1})\cdots\widetilde{O}(k_{n})\\&\times(2\pi)^{4}\delta^{(4)}(k_{1}+k_{2}+\cdots+k_{n})\times\int\frac{d^{4}p}{(2\pi)^{4}}\\&\frac{1}{(p^{2}-m_{0}^{2})[(p+k_{2})^{2}-m_{0}^{2}][(p+k_{2}+k_{3})^{3}-m_{0}^{2}]}\\&\times\cdots
    \frac{1}{[(p+k_{2}+\cdots+k_{n})^{2}-m_{0}^{2}]}.
   \end{split}
\end{equation}
If we drop the terms vanishing in the limit $m_{0}^{2}\ll\tilde{\mu}^{2}$ then we get
\begin{equation}
        \begin{split}
    \Gamma^{n}=&\frac{(-1)^{n}m_{0}^{4}}{32\pi^{2}n(n-1)(n-2)}\int dx_{1}\cdots dx_{n}\frac{d^{4}k_{1}}{(2\pi)^{4}}\cdots\frac{d^{4}k_{n}}{(2\pi)^{4}}\\&\widetilde{O}(k_{1})\cdots\widetilde{O}(k_{n})\delta(\sum_{i=1}^{n}x_{i}-1)\frac{(2\pi)^{4}\delta^{(4)}(k_{1}+\cdots+k_{n})}{(\frac{\Delta}{m_{0}^{2}}+1)^{n-2}}
    \end{split}
\end{equation}
where $\Delta$ is from Feynman parametrization given as
\be
\begin{split}
  \Delta=&\big[x_{2}k_{2}+x_{3}(k_{2}+k_{3})+\cdots+x_{n}(k_{2}+\cdots+k_{n})\big]^{2}\\&-x_{2}k_{2}^{2}-\cdots-x_{n}(k_{2}+\cdots+k_{n})^{2}.
  \end{split}
\ee

To extract the zeroth order term in $\frac{\partial^{2}}{m_{0}^{2}}$ expansion, we replace $(\frac{\Delta}{m_{0}^{2}}+1)^{-(n-2)}$ by 1. The zeroth order term reads
\begin{equation}
    \begin{split}
        \Gamma^{(n)}_{0}=&\frac{(-1)^{n}m_{0}^{4}}{32\pi^{2}n(n-1)(n-2)}\int\frac{d^{4}k_{1}}{(2\pi)^{4}}\cdots\frac{d^{4}k_{n}}{(2\pi)^{4}}\widetilde{O}(k_{1})\cdots\widetilde{O}(k_{n})\\&\times(2\pi)^{4}\delta^{(n)}(k_{1}+\cdots+k_{n})\\&=\frac{(-1)^{n}m_{0}^{4}}{32\pi^{2}n(n-1)(n-2)}\int d^{4}xO(x)^{n}.
    \end{split}
\end{equation}

Summing over all these contributions we get the desired result \eqref{eq:Veff}.
\bibliographystyle{apsrev4-1}
\bibliography{Inflation.bbl}

\end{document}